\magnification=\magstep1
\baselineskip=15pt
\overfullrule=0pt

\def\N{{\cal N}}
\def\O{{\cal O}}
\def\Re{{\rm Re}}

\def\tr{{\rm tr}}

\def\half{{1 \over 2}}
\def\12{{\scriptstyle{1\over2}}}
\def\d{{d \over 2}}

\def\Del2{{\Delta \over 2}}
\def\AdS{{\rm AdS}}

\def\square{{\sqcap \! \! \! \! \sqcup}}

\font\fourteenbf=cmbx10 scaled \magstep2

\input BoxedEPS
\SetRokickiEPSFSpecial  
\HideDisplacementBoxes

\rightline{UCLA/98/TEP/34}
\rightline{MIT-CTP-2795}

\bigskip

\centerline{{\fourteenbf GENERAL SCALAR EXCHANGE IN AdS$_{{\bf
d+1}}$}\footnote{*}{Research supported in part by the National Science
Foundation under grants PHY-95-31023 and  PHY-97-22072.}}

\bigskip
\bigskip
\bigskip

\centerline{{\bf Eric D'Hoker}${}^1$ 
            {\bf and Daniel Z. Freedman} ${}^2$}

\bigskip

\centerline{${}^1$ Department of Physics}
\centerline{University of California, Los Angeles, CA 90095, USA;}

\bigskip

\centerline{${}^2$ Department of Mathematics and Center for Theoretical Physics}
\centerline{Massachusetts Institute of Technology, Cambridge, MA 02139, USA}

\bigskip
\bigskip
\bigskip

\centerline{\bf ABSTRACT}

\bigskip

The scalar field exchange diagram for the correlation function of four scalar
operators is evaluated in anti-de Sitter space, $AdS_{d+1}$. The conformal
dimensions $\Delta_i$, $i=1,\ldots,4$ of the scalar operators and the dimension
$\Delta$ of the exchanged field are arbitrary, constrained only to obey the
unitarity bound. Techniques similar to those developed earlier for gauge
boson exchange are used, but results are generally more complicated. However,
for integer $\Delta_i, \Delta$, the amplitude can be presented as a multiple
derivative of a simple universal function. Results simplify if further
conditions hold, such as the inequalities, $\Delta< \Delta_1+\Delta_3$ or
$\Delta<\Delta_2+\Delta_4$. These conditions are satisfied, with $<$ replaced
by $\le$, in Type IIB supergravity on $AdS_5\times S_5$ because of selection rules
from
$SO(6)$ symmetry. A new form of interaction is suggested for the marginal
case of the inequalities. The short distance asymptotics of the amplitudes
are studied. In the direct channel the leading singular term agrees with the
double operator product expansion. Logarithmic singularities occur at
sub-leading order in the direct channel but at leading order in the crossed
channel. When the inequalities above are violated, there are also $(\log)^2$
singularities in the direct channel.  

\vfill\break

\centerline{\bf I. INTRODUCTION}

\bigskip

Many correlation functions have been evaluated in the study of the AdS/CFT
correspondence [1,2,3]. The 2- and 3-point functions appear to have
non-renormalization properties [4,5] which, although remarkable, suggest that
there is little dynamical content at this level. On the other hand, 4-point
functions are intrinsically more complex in a conformal field theory and are
thus expected to contain more information about the dynamics of the AdS/CFT
correspondence. Studies of their structure have begun both from the AdS side
[6,7,8,9,10,11,12,13,14] and from the viewpoint of the $\N=4$ super-Yang-Mills
boundary theory [14,15,16].

\medskip

A first question is whether the CFT has a simple $t$-channel OPE, so that
the 4-point correlation function admits a convergent power expansion of the form
[10]
$$
\eqalign{
    \langle \O _1 (x_1) & \O _2 (x_2) \O_3 (x_3) \O_4 (x_4) \rangle  \cr
&   = \sum_p {\gamma_{1 3 p} \over (x_1-x_3)^{\Delta_1 + \Delta_3 - 
      \Delta_p}} ~
    {1 \over (x_1-x_2)^{2 \Delta _p}} ~
    {\gamma_{2 4 p} \over (x_2-x_4)^{\Delta_2 + \Delta_4 - \Delta_p}}
\cr}
\eqno (1.1)
$$
containing the contribution of a finite number of primary operators (and
descendents). Explicit calculations of parts of these correlation functions
involving contact interactions [9] or gauge boson exchange [11], show that
logarithmic singularities occur as well as the expected leading powers. Thus far,
no full correlator of chiral primary operators has been computed, so it is 
not yet known whether logarithms will cancel or survive when all contributions
are assembled. Logarithms need not be inconsistent with (1.1) and it as
recently conjectured [17] that they are the effects of anomalous dimensions of
double trace operators in the boundary theory.

\medskip

A second question is whether the 4-point function, presumably upon including
full towers of string excitation states, exhibits crossing symmetry, i.e.
duality. Crossing symmetry holds universally for the 4-point functions of CFT
in 2 space-time dimensions, and plays a crucial role there. To investigate this
issue, one needs to be able to study the OPE for exchange states of arbitrary
masses and spins which are not necessarily chiral primary or descendents
thereof, but may be non-chiral primary.

\medskip

In the present paper, we consider the correlation function of 4 scalar
operators $\O_{\Delta _i}$ of arbitrary scaling dimension $\Delta _i$,
$i=1,\cdots ,4$, and study the corresponding Witten diagram in which a scalar
field of arbitrary dimension $\Delta$ is exchanged. (See Fig.~1.)
We assume throughout that all scaling dimensions obey the unitarity bound
$\Delta _i \geq d/2$ and $\Delta \geq d/2$.  We generalize the expansion and
resummation techniques developed in [11] to evaluate the integrals over the
interaction points $z$ and $w$ in Fig.~1.

\medskip

For general integer values of $d$, $\Delta _i$ and $\Delta$, the amplitude is
more complicated than in [11], and is expressed as a double
infinite series. If the combination
$$
\delta = \half (\Delta _1 + \Delta _2 + \Delta _3 + \Delta _4)
\eqno (1.2)
$$
is an integer, then one can introduce a simple universal function $I(\tau;s,t)$
of a parameter $\tau$ and two conformal invariant combinations $s$ and $t$
of the four points $x_i$, but independent of $\Delta$, $\Delta _i$ and $d$. The
correlator can then be expressed as a multiple derivative with respect to $x_i$
of the integral in $\tau$ of a product of $I(\tau;s,t)$ and a ``spectral weight
function" $\rho (\tau)$, which in turn is defined by the double infinite
series above.

\medskip

If one further condition is satisfied, namely that $\delta - \Delta _3 -d/2$ is
a non-negative integer, then one of the infinite series above truncates to a
finite sum. If the second condition $\Delta < \Delta _1 + \Delta _3$ holds,
then the second infinite series also truncates to a finite sum. These are very
significant simplifications, and it is quite remarkable to observe that all
three conditions above hold in the application of the AdS/CFT correspondence
of greatest current interest, namely Type IIB superstring theory on
$AdS_5\times S_5$, as a consequence of the $SO(6) \sim SU(4)/Z_2$ symmetry
of the sphere $S_5$. 

\medskip

These truncations take place because the Kaluza-Klein scalar fields on $\AdS_5\times
S_5$ involve [18,4] $S_5$ scalar spherical harmonics $Y^k$ (and generalizations) where
$k$ denotes a $k$-fold product of the vector 6-dimensional representation of $SO(6)$.
One  may define the parity of $Y^k$ to be $(-)^k$, a quantity related to the $SU(4)$
quadrality of the representation. Invariant interactions necessarily involve the
contraction of an {\it even} number of vector indices, so the net parity of each vertex
in the diagram of  Fig.~1 must be even. It is also the case [18] that scale dimensions
are  linearly related to $k$ such that $(-)^\Delta = (-)^k$ for every scalar field. The
parity selection rule then guarantees that $\Delta + \Delta _1 +\Delta _3$ and $\Delta +
\Delta _2 +\Delta _4$ are both even, so the first condition of the previous paragraph
automatically holds (since $d=4$).

\medskip

The basic reason why an inequality of the form $\Delta < \Delta _1 + \Delta _3$
might hold in the $AdS_5 \times S_5$ theory is that the tensor product
$Y^{k_1} Y^{k_3}$ of two harmonics contains only representations which
satisfy $|k_1 - k_3| \leq k \leq k_1 + k_3$. To use this information in a more
precise way, we note that each infinite Kaluza-Klein tower of scalar
fields in [18] is characterized by a non-negative integer $s$, such that
$\Delta = 2s +k$ for all fields in the tower. The cases $s=0,1,2,3$ occur, and
$s=0$ holds for the Kaluza-Klein tower corresponding to the chiral primary
operators $\tr X^k$ of the boundary super Yang-Mills theory. For this family
of scalars, the Clebsch-Gordon limit above immediately gives $\Delta \leq 
\Delta _1 + \Delta _3$. (The marginal case requires special discussion;
see Appendix~2.) For cubic vertices involving other Kaluza-Klein families a precise
statement requires further study.

\medskip

In Sec.~2 below, we discuss the propagators and vertices which are the 
ingredients of the diagram of Fig.~1. In Sec.~3, we define the amplitudes 
precisely and carry out the $z$- and $w$-integrals of Fig.~1, and derive a
two-dimensional parametric integral representation of the amplitude for general
$\Delta _i$, $\Delta$ and $d$. In Sec.~4, we discuss the simplifications that
take place when special conditions on $\Delta _i$, $\Delta$ and $d$ are
satisfied and we show how to present the amplitude in terms of a single
parametric integral with a universal function. In Sec.~5, we first apply a simple
method to obtain the leading singularity in the direct channel, as $x_2 - x_4
\to 0$ and then use general methods to derive the full asymptotic behavior of the
amplitude as $x_2 - x_4\to 0$ and $x_2 - x_3 \to 0$. In Sec.~6, we
present brief conclusions. In the Appendix we propose a new cubic interaction for
fields satisfying $\Delta = \Delta_1 + \Delta_3$.

\midinsert
$$
\BoxedEPSF{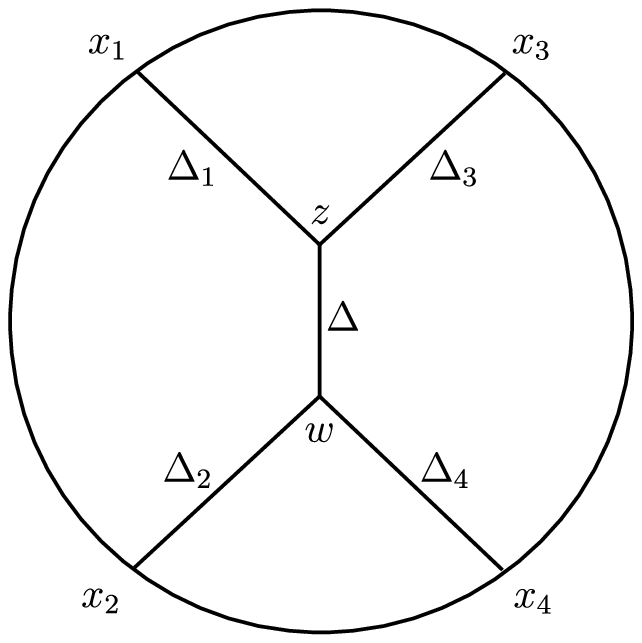 scaled 666}  
$$
\centerline{Figure 1}
\endinsert

\bigbreak

\centerline{{\bf 2. THE SCALAR PROPAGATOR}}

\bigskip

We work on the Euclidean continuation of $\AdS_{d+1}$ viewed as the upper half
space in $z_\mu \in {\bf R}^{d+1}$, with $z_0 >0$, and metric $g_{\mu \nu}$ of
constant negative curvature $R=-d(d+1)$, given by
$$  
  ds^2 = \sum _{\mu, \nu=0} ^ d g_{\mu \nu} dz_\mu dz_\nu
       ={1 \over z_0^2} (dz^2_0 + \sum ^d_{i=1} \, dz^2_i)
\, .
\eqno (2.1)
$$
It is well-known that AdS-invariant functions, such as scalar propagators, are
simply expressed [19] as functions of (the  chordal distance)  $u$, defined by
$$
  u  = {(z-w)^2 \over 2z_0 w_0}
\qquad \qquad
(z-w)^2 = \delta_{\mu \nu}(z-w)_{\mu} (z-w)_{\nu}
\eqno (2.2)
$$

\medskip

We shall consider a general field theory of real scalars $\Phi _\Delta$
with masses 
$$
m^2 = \Delta (\Delta -d) \, ,
\eqno (2.3)
$$
which are sources for the  boundary operators $\O _\Delta (x)$ of
dimensions $\Delta$ obeying the unitarity bound $\Delta \geq d/2$. For the
sake of simplicity, we shall assume that the couplings of the scalar fields
are non-derivative and trilinear. Amplitudes with derivative couplings may be
reduced to amplitudes with non-derivative couplings only, as shown in [8,9].
Besides trilinear couplings, only quadrilinear couplings might be required in
evaluating the four point function, and contact contributions with these
interactions were evaluated in [9]. The action then takes the form
$$
S 
  = \sum _{\Delta } \int d^{d+1}z \, \sqrt{g} 
      \left[  \Phi _\Delta ( -\square _g  + m^2) \Phi _\Delta  \right] 
 +  \sum _{\Delta, \Delta', \Delta ''} \gamma _{\Delta \Delta ' \Delta ''}
      \Phi _{\Delta } \Phi _{\Delta '} \Phi _{\Delta ''},
\eqno (2.4)
$$
where $\square _g = D^\mu D_\mu$ is the scalar Laplacian with metric $g$ on
AdS, and $\gamma _{\Delta \Delta ' \Delta ''}$ are trilinear couplings, which
we set to unity.

\medskip

The scalar bulk to boundary propagator for dimension $\Delta$ is given by [3]
$$
   K_\Delta (z_0, \vec{z}, \vec{x}) 
       = C_\Delta \left( 
                    {z_0 \over z^2_0 + \left(\vec{z} - \vec{x} \right)^2}
                  \right)^\Delta
 \eqno (2.5)
$$
with the following normalizations for the constant prefactors [20],
$$
C_\Delta =  {\Gamma (\Delta) \over \pi^{d/2} \Gamma (\Delta - d/2)}
\quad {\rm for} \quad \Delta > d/2, \qquad \qquad 
C_{d/2} =  { \Gamma (d/2) \over 2 \pi^{d/2}} \, . 
\eqno (2.6)
$$

The scalar bulk-to-bulk propagator for dimension $\Delta$ was obtained in [19], 
$$
\eqalign{
G_\Delta (u)
& =
  \tilde C _\Delta (2u^{-1} )^\Delta 
  F \bigl (\Delta , \Delta -\d +\half; 2\Delta -d +1; -2u^{-1} \bigr ) \cr
  \tilde C _\Delta & = {\Gamma (\Delta) \Gamma (\Delta -\d -\half)
                        \over  (4\pi)^{(d+1)/2} \Gamma (2 \Delta -d+1)}
\cr}
\eqno (2.7)
$$
Here, $F$ is the standard hypergeometric function ${}_2F _1$.
It will be very convenient for later use to make a quadratic transformation of
the hypergeometric function [21], and to recast it in terms of the variable 
$$
\xi \equiv {1 \over 1+u} = { 2 z_0 w_0 \over \left( z^2_0 + w^2_0 + (\vec{z} -
  \vec{w})^2 \right)} \, .
\eqno (2.8)
$$
We obtain
$$
G_\Delta (u) 
 =
   2^\Delta \tilde C _\Delta \xi ^\Delta 
   F \big (\Del2, \Del2 + \half; \Delta -\d +1;\xi ^2 \big )
\, .
\eqno (2.9)
$$ 
Since from (2.8) it is clear that $ |\xi| \leq 1$, it follows immediately
that the form (2.9) of the scalar propagator will have a uniformly convergent
expansion in powers of $\xi $. Thus, the summation of the series expansion of
$F$ in powers of $\xi $ and any convergent integration of $G(u)$ may be freely
interchanged. This property will be very useful later on. 

\bigskip
\bigbreak

\centerline{{\bf 3. SCALAR EXCHANGE INTEGRALS}}

\bigskip

We factor out the normalization constants and define the
amplitude to be studied by
$$
  \left\langle \O_{\Delta _1} (x_1) \O_{\Delta _2} (x_2)
     \O_{\Delta _3} (x_3) \O_{\Delta _4} (x_4) \right\rangle 
   = 2^\Delta \tilde C_\Delta C_{\Delta _1} C_{\Delta _2} C_{\Delta _3} 
C_{\Delta _4}
    A (x_1, x_2, x_3, x_4)
\eqno (3.1a)
$$
with
$$
\eqalign{
    A (x_i) = 
      \int {d^{d+1}z \over z^{d+1}_0} \int {d^{d+1}w \over w^{d+1}_0} &\,
\xi ^\Delta
  F\big ({\Delta \over 2}, {\Delta \over 2} + \half; \Delta -\d +1;\xi ^2 \big) 
      \left({z_0 \over (z - x_1)^2} \right)^{\Delta _1}  \cr
   &  \cdot \left({z_0 \over (z - x_3)^2} \right)^{\Delta _3}
      \left({w_0 \over (w - x_2)^2} \right)^{\Delta _2}
      \left({w_0 \over (w - x_4)^2} \right)^{\Delta _4}
\cr}
\eqno (3.1b)
$$
The convergence conditions of this integral, for $x_i$ fixed and separated from
one another, are easily obtained by inspection. First, assuming the unitarity
bound on all dimensions $\Delta, \ \Delta _i \geq \d$, it is clear that the only
divergences arise from when one or two interaction points approach one of the
boundary points $x_i$. Convergence when one of the interaction points
approaches any one of the boundary points is guaranteed by the conditions
$$
|\Delta _1 - \Delta _3 | < \Delta
\qquad \qquad
|\Delta _2 - \Delta _4 | < \Delta\, ,
\eqno (3.2a)
$$
while convergence when both interactions points approach any one of the
boundary points requires
$$
2\Delta _i < \Delta _1 + \Delta _2 + \Delta _3 + \Delta _4 \, ,
\qquad i=1,2,3,4\, .
\eqno (3.2b)
$$
Conditions (3.2a) are familiar from the case of the three point functions [20].
Throughout, we shall assume that the above convergence conditions hold.

\medskip

The first step [9,11,20] is to simplify the integral by setting $x_1 = 0$,
and by changing integration variables using the inversion isometry of AdS,
namely $z_\mu = z'_\mu / (z')^2$ and $w_\mu = w'_\mu / (w')^2$,
with boundary points $x_2,x_3,x_4$ referred to their inverses by $x_i =
x'_i / (x_i ')^2$. Once this inversion is carried out and $x'_1=\infty$, the
reduced function is invariant under simultaneous translations of $x'_2$, $x'_3$
and $x'_4$, and thus only depends upon $x \equiv x'_4 - x'_3$ and $y\equiv
x'_2 - x'_3$. The net result of the inversion and change of variables is
$$
  A (x_1, x_2, x_3, x_4) =   
         |x'_2|^{2 \Delta_2} |x'_3|^{2 \Delta _3 } |x'_4|^{2 \Delta _4}
        B (x, y)\, .
\eqno (3.3)
$$
The function $B(x,y)$ is then easily found from (3.1), and is given by
$$
\eqalignno{
B (x,y) 
  =& \int d^{d+1} w  
         \ {w_0^{\Delta _2 + \Delta _4 -d-1} 
            \over (w-x)^{2 \Delta _4} (w-y)^{2 \Delta _2}}\  R(w) & (3.4a) \cr
&&\cr
R (w) 
  =& \int d^{d+1} z 
     \ {z_0^{\Delta _1 + \Delta _3 -d - 1} \over (z_0 ^2 + \vec{z} ^2) 
           ^{\Delta _3}} \ \xi ^\Delta\ 
  F\big (\Del2 , \Del2 + \half; \Delta -\d +1;\xi ^2 \big) 
& (3.4b) \cr}
$$
where we have dropped the primes on the integration variables $w, z$, and $\xi $
continues to be defined by (2.8), as a function of $z$ and $w$.

\bigskip

\noindent
{\bf a. Integrals over the interaction point $z$}

\medskip

We now evaluate the $z$-integrals, which define $R(w)$, by using the
uniformly convergent expansion of the hypergeometric function in (3.4b) in
powers of $\xi $.
$$
\eqalignno{
R(w) 
  & = \sum _{k=0} ^\infty 
    {\Gamma (k + \Del2) \Gamma (k + \Del2 + \half) \Gamma (\Delta -\d +1)
     \over
     \Gamma (\Del2) \Gamma (\Del2 + \half) \Gamma (k +\Delta -\d +1) k! }
   \ (2 w_0 )^{\Delta + 2k} \ R_k(w) 
& (3.5a) \cr
&&\cr
R_k (w)
  & = \int _0 ^\infty dz_0 \int d^d z \
     {z_0 ^{\Delta + \Delta _1 + \Delta _3 -d-1 +2k}
      \over
      \big [ z_0 ^2 + w_0 ^2 + (\vec{z} - \vec{w})^2\big ] ^{\Delta +2k} } \
     {1 \over (z_0 ^2 + \vec{z} ^2) ^{\Delta _3}}
& (3.5b) \cr}      
$$
The spatial integral in (3.5b) is convergent for $\Delta _1,\ \Delta _3 \geq
\d$ and may be carried out by introducing a Feynman parameter $\alpha$, yielding
$$
  {\pi^{d/2} \Gamma (\Delta + \Delta _3 +2k -\d) 
   \over \Gamma (\Delta + 2k) \Gamma (\Delta _3)}
  \int^1_0 \!\! d \alpha \
  { \alpha^{\Delta + 2k-1} (1 -
    \alpha)^{\Delta _3 -1} \over \big [ z^2_0 + \alpha (1 - \alpha) 
      \vec{w}^2 + \alpha w^2_0 \big ]^{\Delta + \Delta _3 +2k -\d}}
$$
The $z_0$ integral is then straightforward, and we find 
$$
R_k(w) = 
  {\pi^{d/2} \Gamma (k + a)
             \Gamma (k + b) 
   \over 2 \Gamma (\Delta + 2k) \Gamma (\Delta _3)}
 \int^1_0 d\alpha \, 
  { \alpha^{k +\Delta - b -1} (1 -
    \alpha)^{\Delta _3 -1} \over \big [ (1 - \alpha) 
      \vec{w}^2 + w^2_0 \big ]^{k + b}} \, .
\eqno (3.6)
$$
Here, we define the following abbreviations ($c$ will enter shortly)
$$
\eqalign{
a & \equiv \half(\Delta + \Delta _1 + \Delta _3) -\d \cr
b & \equiv \half(\Delta - \Delta _1 + \Delta _3) \cr
c & \equiv \Delta -\d +1 \, . \cr}
\eqno (3.7)
$$
Actually, this integration will be convergent for all $k\geq 0$ only for $b >0$,
but this is guaranteed by (3.2a).

\medskip

Putting together this result and the summation over $k$ at first sight yields
individual terms that have four $k$-dependent $\Gamma $ factors in
the numerator, and three others in the denominator. Happily however, the
following special ratio always occurs and may be simplified with the help of the
doubling formula of the Euler $\Gamma$-function.
$$
{\Gamma (k + \Del2) \Gamma (k + \Del2 + \half) \Gamma (\Delta) 
 \over 
\Gamma (\Del2) \Gamma (\Del2 + \half) \Gamma (2k + \Delta)} = 
{1 \over 2^{2k}} \, .
\eqno (3.8)
$$
Using this result, the summation is now over terms with two $k$-dependent
$\Gamma $ functions in the numerator, one in the denominator, as well
as a factor of $k!$ in the denominator, times a $k$-power of a composite
variable 
$$
\gamma \equiv { \alpha w_0 ^2 \over w_0 ^2 + (1 -\alpha ) \vec{w}^2} \, .
\eqno (3.9)
$$ 
This expression thus resums to a standard hypergeometric function, and we may
recast $R(w)$ in the form
$$
R(w) = 2^{\Delta -1} \pi ^{\d} w_0 ^{\Delta _1 - \Delta _3}
  {  \Gamma (a) \Gamma (b) \over \Gamma (\Delta _3) \Gamma (\Delta)}
  \int _0 ^1 d\alpha \ \alpha ^{\Delta _1 - \Delta _3 -1} 
  (1-\alpha)^{\Delta _3-1}
   \gamma ^b F(a,b;c;\gamma) 
\eqno (3.10)
$$
where the combinations $a$, $b$, $c$ were defined in (3.7), and $\gamma$ is given
by (3.9) in terms of $\alpha$. Notice that as $\alpha$ runs from 0 to 1, so
does $\gamma$, and the integral may be rewritten in terms of an integration over
$\gamma$.
$$
R(w) = 2^{\Delta -1} \pi ^{\d} 
   {w_0 ^{\Delta _1 + \Delta _3}\over w^{2 \Delta _3 - 2 \Delta _1}}
  {  \Gamma (a) \Gamma (b) \over \Gamma (\Delta _3) \Gamma (\Delta)}
  \int _0 ^1 d\gamma \, {\gamma ^{\Delta - b -1} (1-\gamma)^{\Delta _3-1} 
   \over \big [ (1-\gamma) w_0 ^2 + \gamma w^2 \big ]^{\Delta _1}} 
     F(a,b;c;\gamma) \, .  
\eqno (3.11)
$$
In principle, the $\gamma$-integration may be carried out in terms of a
generalized hypergeometric function ${}_4F_3$, but expressions of this type
tend to be less useful than explicit series expansions, which we shall use
instead. 

\bigskip

\noindent
{\bf b. Integrals over the interaction point $w$}

\medskip

Following (3.4a), the reduced amplitude $B(x,y)$, which gives the full amplitude
via (3.3), is obtained from an integration over $w$. Interchanging the
convergent $\gamma$ and $w$-integrations, we have
$$
B(x,y) =  2^{\Delta -1} \pi ^{\d} 
  {  \Gamma (a) \Gamma (b) \over \Gamma (\Delta _3) \Gamma (\Delta)}
  \int _0 ^1 \! d\gamma \, \gamma ^{\Delta - b -1}
(1-\gamma)^{\Delta _3-1} \,
  F(a,b;c;\gamma) \, W(x,y;\gamma)
\eqno (3.12)
$$
where the function $W(x,y;\gamma)$ is defined by
$$
W(x,y;\gamma)
   = \int d^{d+1} w  
         \ {w_0^{\Delta _1 + \Delta _2 + \Delta _3 + \Delta _4 -d-1} 
            \over (w-x)^{2 \Delta _4} (w-y)^{2 \Delta _2}}\
            { w^{2\Delta _1 -2 \Delta _3} \over \big [ (1-\gamma) w_0^2 +
\gamma w^2 \big ] ^{\Delta _1}} \, .
\eqno (3.13)
$$
The $w$-integral is convergent for all $x$ and $y$, and $0\leq \gamma
\leq 1$. To evaluate it, it is most convenient to proceed by Taylor
series expansion in powers of $w_0^2 /w^2$ under the integration of the
denominator $(1-\gamma)w_0^2 +\gamma w^2$.  However, such an expansion is
convergent only when $\half <\gamma \leq 1$. We shall evaluate the
function $W$ in this range of $\gamma$ (actually, $\half < \Re (\gamma) 
\leq 1$) and then analytically continue to the full range of $\gamma$. We
shall be able to see explicitly that the analytic continuation encounters
no singularities, and is in fact automatic.

\medskip

Thus, for $\half < \gamma \leq 1$, we evaluate $W$ by the following Taylor
series expansion
$$
\eqalignno{
W(x,y;\gamma) & = \sum  _{k=0} ^\infty 
                  {\Gamma (k + \Delta _1) \over  \Gamma (\Delta _1)\ k!}
                   (\gamma -1)^k \gamma ^{-k - \Delta _1} W_k (x,y)
& (3.14a) \cr
W_k(x,y) & = \int \! d^{d+1}w 
              {w_0^{2k+\Delta _1 + \Delta _2 + \Delta _3 + \Delta _4 -d-1} 
              \over (w - x)^{ 2\Delta _4} \
                    (w- y)^{2 \Delta _2}\
            w ^{2\Delta _3 + 2k } } 
& (3.14b) \cr}
$$  
The integrals $W_k(x,y)$ can now be done by the following standard steps:
  (a) combine $w^2$ and $(w-y)^2$ denominators with Feynman parameter
      $\alpha$,
  (b) combine the composite denominator from the previous step with the
      $(w-x)^2$
      denominator using Feynman parameter $\beta$,
  (c) carry out the $d^dw$ integral,
  (d) do the $dw_0$ integral. 
We suppress details and directly give the result in terms of the following
combination of dimensions that will enter throughout
$$
\delta \equiv \half (\Delta _1 + \Delta _2 + \Delta _3 + \Delta _4)\, .
\eqno (3.15)
$$
After several simplifications, we get
$$
\eqalign{
    W_k(x,y) &= {\pi^{d/2} \Gamma (k+\delta -\d) \Gamma 
                  (\delta - \Delta _1) \over 
                2 \Gamma (\Delta _2) \Gamma (\Delta _4) \Gamma
                  (k + \Delta _3)}
              \cr
        & \quad {} \cdot
           \int^1_0 d \alpha
           \int^1_0 d \beta \
           {\alpha^{\Delta_2 - 1} (1 - \alpha)^{k +\Delta _3 -1}
                \beta^{\Delta _4-1} (1-\beta)^{k+\delta - \Delta _4 -1} 
             \over \left [ \beta (x - \alpha y)^2 + \alpha (1 -
                 \alpha) y^2 \right]^
               {\delta - \Delta _1 }}
\cr}
\eqno (3.16)
$$
Notice that (3.14) was convergent only when $\delta - \Delta _1 >0$, but this
is precisely the convergence condition (3.2b).

Assembling this result for $W_k(x,y)$ in the sum (3.14a), we notice that the
summation over $k$ becomes proportional to a hypergeometric function, so that
the function $W(x,y;\gamma)$ takes on the form
$$
\eqalign{
W(x,y;\gamma)
    = {\pi^{d/2} \Gamma (\delta -\d) \Gamma 
                  (\delta - \Delta _1) \over 
                2 \Gamma (\Delta _2) \Gamma (\Delta _3) \Gamma (\Delta _4) }
& \gamma ^{-\Delta _1}
           \int^1_0 \! d \alpha
           \int^1_0 \! d \beta \ F(\Delta _1, \delta -\d; \Delta _3;
  -\sigma) \cr
& 
         \cdot  {\alpha^{\Delta_2 - 1} (1 - \alpha)^{\Delta _3 -1}
                \beta^{\Delta _4-1} (1-\beta)^{\delta - \Delta _4 -1} 
             \over \left [ \beta (x - \alpha y)^2 + \alpha (1 -
                 \alpha) y^2 \right]^
               {\delta - \Delta _1 }} 
\cr}
\eqno (3.17)
$$
where $\sigma$ is defined by
$$
\sigma = (1-\alpha) (1-\beta )(1-\gamma)/\gamma
\eqno (3.18)
$$
Now, originally, this expression was valid only for $\half < \gamma \leq 1$,
and the function $W(x,y;\gamma)$ for the full range of integration of $\gamma$
was to be defined by analytic continuation in $\gamma$. The dependence of
$\sigma$ on $\gamma$ makes it clear that for any value of $\gamma$ in the
integration range, we have $\sigma \geq 0$, so that the hypergeometric function
is well-defined for all $0\leq \gamma \leq 1$, and the only singularity in
(3.17) is the power prefactor $\gamma ^{-\Delta _1}$. Thus, the analytic
continuation in $\gamma$ is straightforward, and (3.17) is the correct
expression for $W(x,y;\gamma)$ for {\it all values } of $0<\gamma \leq 1$.

\bigskip

\noindent
{\bf c. Combining all integrals}

\medskip

The expression in (3.17) may be recast in a more useful way by performing 
an analytic continuation [22] on the hypergeometric function, in such a way
that its new argument is brought into the standard range $[0,1]$. The
transformation takes the form
$$
F(\Delta _1, \delta -\d;\Delta _3;-\sigma)
   = (1-\eta)^{\Delta _1} F (\Delta _1, \Delta _3 -\delta +\d; \Delta
_3;\eta)
\eqno (3.19)
$$
where $\sigma$ was defined in (3.18) and $\eta = \sigma /(1+\sigma)$, with
$0\leq \eta \leq 1$. 

\medskip

The best formulas for deriving asymptotics are those where the
integration range is scale invariant, so we change integration variables for
the Feynman parameters $\alpha$ and $\beta$ to variables $u$ and
$v$,\footnote{*}{This change of variables was carried out in 3 stages in [11]
to obtain (3.29) of [11].} and express also $\eta$ in terms of $u$ and
$v$ 
$$
\alpha = {1 \over 1 + u}, 
\qquad  
\beta = {u \over uv + u + v}
\qquad \qquad
\eta = {(1-\gamma) uv \over uv + \gamma (u+v) }
\eqno (3.20)
$$
After a number of simplifications, the final expression for $W$ is given by
$$
\eqalign{
W(x,y;\gamma)
  = {\pi^{d/2} \Gamma (\delta -\d) \Gamma 
                  (\delta - \Delta _1) \over 
                2 \Gamma (\Delta _2) \Gamma (\Delta _3) \Gamma (\Delta _4) } 
\int _0 ^\infty \! du   \int _0 ^\infty \! dv &  \
F(\Delta _1, \Delta _3 -\delta +\d; \Delta _3; \eta)\cr  
\cdot { 1
  \over
\big [ uv + \gamma (u+v) \big ] ^{\Delta _1} } 
&
\cdot {  u^{\delta - \Delta _2 -1} v^{\delta - \Delta _4 -1}
\over 
\big [ u x^2 + v y^2 + (x-y)^2 \big ] ^{\delta - \Delta _1} }.
\cr}
\eqno (3.21)
$$

\medskip

Finally, the hypergeometric function $F(a,b;c;\gamma)$ in (3.12) may be
transformed into a more useful form by the following formula
$$
F(a,b;c;\gamma) = (1-\gamma) ^{c-a-b} F(c-a,c-b;c;\gamma)\, .
\eqno (3.22)
$$
The advantage of the form on the right hand side is two-fold. First, when
$\Delta < \Delta _1 + \Delta _3$, this form will truncate to a finite
polynomial, in analogy with the case of the photon exchange [11]. Second, for
the parameter values of $a,b,c$ of (3.7), the series expansion of
$F(a,b;c;\gamma)$ is not  uniformly convergent, while that of
$F(c-a,c-b;c;\gamma)$ is uniformly convergent for $\gamma \in [0,1]$. Uniform
convergence will allow us to systematically expand $F$ in a power series and
treat the series term by term. 

\medskip

We are now ready to combine all contributions and present a completely general
formula for the scalar exchange four point function in arbitrary dimension. The
reduced amplitude $B(x,y)$ of (3.3) is given by
$$
\eqalignno{
B(x,y) & =
   {\pi^d \Gamma (\delta -\d) \Gamma (\delta - \Delta _1) 
              \Gamma (a) \Gamma (b)
\over 
   2^{2-\Delta} \Gamma (\Delta) \Gamma (\Delta _2) \Gamma (\Delta _3) ^2  
                \Gamma (\Delta _4) } \ B_R(x,y) 
& (3.23a) \cr
B_R(x,y) & =  
    \int _0 ^\infty \!\!\! du \int _0 ^\infty \!\!\! dv \,
      {u^{\delta -\Delta _1 -\Delta _2 -1} v^{\delta -\Delta _1 -\Delta _4 -1}
         \over
      \big [ ux^2 + vy^2 +(x-y)^2 \big ] ^{\delta - \Delta _1} }\ 
      \rho \left ( {uv \over u+v +uv} \right ) 
& (3.23b) \cr}
$$
with the ``spectral density" function $\rho $ given by
$$
\rho (\tau)  = \tau ^{\Delta _1}
\int _0 ^1 \!\! d \gamma \
       \gamma ^{\Delta - b -1}
       F(c-a,c-b;c;\gamma) \ {F(\Delta _1, \Delta _3 -\delta +\d; \Delta _3;\eta)
       \over
       ( \gamma + \tau - \gamma \tau ) ^{\Delta _1} }\, . 
\eqno (3.24)
$$ 
The function $\rho $ depends only on the combination
$uv/(u+v+uv)$ and on the dimensions. Here, $\eta$ of (3.20) is expressed in
terms of $\tau$ and $\gamma$ by the relation
$$
\eta = {(1-\gamma) \tau \over \gamma + \tau -\gamma \tau }\, .
\eqno (3.25)
$$ 
Notice that the denominator of $\eta$ is the same as the one occurring in
(3.24b), and that the numerator factorizes; this will be very useful in the
next section.

\bigskip

\noindent
{\bf d. Infinite series expansions}

\medskip

It will be advantageous to express the second hypergeometric function in
(3.23b) and (3.24b) in terms of an infinite sum. In particular, since $\eta$ is a
composite variable that depends on all $u$, $v$ and $\gamma$, this expansion
will better allow us to carry out the integrations. Also, as we shall
demonstrate in the subsequent section, under certain mild conditions
(which are always satisfied in the case of $\AdS_5\times S_5$) on the dimensions
$d$ and $\Delta _i$, this infinite series truncates to a sum over a finite number
of terms. We have
$$
\rho (\tau)  =
   \sum _{\ell =0} ^\infty 
     { \Gamma (\ell + \Delta _1) \Gamma (\ell + \Delta _3 - \delta +\d) \Gamma
       (\Delta _3) \over
      \Gamma (\Delta _1) \Gamma (\Delta _3 - \delta + \d) 
      \Gamma (\ell + \Delta _3) \Gamma (\ell +1) } \ \rho _\ell (\tau)\, , 
\eqno (3.26) 
$$
with the functions $\rho _\ell$ given by
$$
\rho _\ell (\tau)  = \tau ^{\ell +\Delta _1}
\int _0 ^1 \!\! d \gamma \   \gamma ^{\Delta - b -1}
(1-\gamma)^\ell
       \ {F(c-a,c-b;c;\gamma)
       \over
       ( \gamma + \tau - \gamma \tau ) ^{\Delta _1 +\ell} } \, .
\eqno (3.27)
$$
A simple relation gives $\rho _\ell$ in terms of $\rho _0$,
$$
\rho _\ell (\tau) 
=
(-)^\ell \tau ^{\ell +\Delta _1} {\Gamma (\Delta _1) \over \Gamma (\Delta _1
+\ell)} {\partial ^\ell \over \partial \tau ^\ell} \ \{ \tau ^{-\Delta _1} \rho
_0 (\tau) \} \, .
\eqno (3.28)
$$
It remains to study $\rho _0$.

\medskip

Using the property of uniform convergence of the series expansion of the 
hypergeometric function in powers of $\gamma$ for $1+a+b-c=\Delta _3 \geq 2$
throughout $\gamma \in [0,1]$, we obtain
$$
\rho _\ell (\tau) 
 = 
\sum _{k=0} ^\infty 
  {\Gamma (c-a+k) \Gamma (c-b+k) \Gamma (c) 
     \over
   \Gamma (c-a) \Gamma (c-b) \Gamma (c+k) \Gamma (k+1)}
\ \rho _\ell ^{(k)} (\tau)
\eqno (3.29) 
$$  
The function $\rho _\ell ^{(k)} (\tau)$ is clearly proportional to a
hypergeometric function, given by
$$
\rho _\ell ^{(k)} (\tau) 
 = \tau ^{\Delta _1+\ell}
{\Gamma (\ell +1) \Gamma (\Delta -b+k) \over \Gamma (\Delta -b+k+\ell+1)}
F(\ell+1,\ell + \Delta _1;\Delta -b+k+\ell+1;1-\tau)\,
\eqno (3.30) 
$$
From the form (3.30) it is clear that $\rho _\ell ^{(k)}(\tau)$
is a (complicated) elementary function, since the hypergeometric function for
the arguments of (3.30) is degenerate and reduces to a combination of rational
and logarithmic contributions [21].

\medskip

For $|\tau|<1$, the hypergeometric function of (3.30) also admits an
absolutely convergent series expansion in $1-\tau$, which may be combined with
(3.29) into a double series. Exchanging the orders of summation, and
generalizing to the case of arbitrary value of $\ell$, we get
$$
\rho _\ell (\tau)  =
  \tau ^{\Delta _1 +\ell} \sum _{p=0} ^\infty Q_{p+\ell} (1-\tau)^p
{\Gamma (\Delta _1) \Gamma (\ell +p+1) \over \Gamma (\Delta _1 +\ell) \ p!}\, . 
\eqno (3.31) 
$$
The coefficients $Q_p$ depend only upon the dimensions and are given by
$$
\eqalignno{
Q_p & \equiv
  \sum _{k=0} ^\infty 
   {\Gamma (c-a+k) \Gamma (c-b+k) \Gamma (c) \Gamma (\Delta -b+k)         
    \Gamma (p+\Delta _1) 
    \over 
    \Gamma (c-a) \Gamma (c-b) \Gamma (c+k) \Gamma (k+1) 
    \Gamma (\Delta -b+k+p+1) \Gamma (\Delta _1)
}\, ,
  & (3.32a) \cr
& = {\Gamma (\Delta -b) \Gamma (p+\Delta _1) \over \Gamma (\Delta -b +p+1)
\Gamma (\Delta _1)}\ {}_3F_2(c-a,c-b,\Delta -b; c, \Delta -b +p+1;1) 
& (3.32b)
\cr}
$$
which is convergent for all $p\geq 0$, and for large $p$ behaves like
$Q_p \sim p^{{\rm Min} (0,a-c)}$. The function ${}_3F_2$ is the standard
generalized hypergeometric function. 

\bigskip
\bigbreak

\centerline{\bf 4. SPECIAL CONDITIONS ON $\Delta_i, \Delta$}

\bigskip

Henceforth, we shall restrict attention to the case of four point functions in
which the dimensions $\Delta$ and $\Delta _i$, $i=1,\ldots ,4$ are  integers
and larger than or equal to the unitarity bound $\d$. Under these restrictions,
$a$, $b$, $c$ and $\delta$ are half integers. We shall now show that if
certain combinations of these numbers are actually integers (possibly of
definite sign), considerable simplifications occur in (3.23), (3.26) and (3.29),
and we shall work out the simplified expressions explicitly in those cases. 

The following cases below are mutually independent from one another. 

\medskip

\noindent
{\bf a. Integer valued $\delta -\d - \Delta _3 \geq 0$ : Truncating the
$\ell$-series.}

\smallskip

When $\delta -\d - \Delta _3$ is integer and $\geq 0$, it is straightforward to
see that the infinite series over $\ell$ in (3.26) truncates to a finite sum,
with $\delta -\d - \Delta _3 +1$ terms. For the case of most
physical urgency $\AdS _5 \times S_5$ with $d=4$, the remaining condition that
$\delta$ is integer and that $\delta > \Delta _3 +1$ is automatically satisfied
thanks to the $R$-parity considerations given in the introduction, and the
convergence conditions of (3.2b). 

\medskip

\noindent
{\bf b. The case $\Delta < \Delta _1 + \Delta _3$ : Truncating the
$k$-series}

\smallskip

When $c-a = 1+ (\Delta -\Delta _1 -\Delta _3)/2  \leq 0$, the infinite series
expansion of the spectral density $\rho _0$ and thus of $\rho _\ell$ 
truncates to a finite sum, with $(\Delta _1 + \Delta _3 - \Delta +1)/2$ terms.
In other words, this is the case when a triangle inequality holds
$$
\Delta < \Delta _1 + \Delta _3
\eqno (4.1)
$$
which is analogous to (3.2a). More generally, if either $\Delta _1 + \Delta _3
>\Delta $ or $\Delta _2 + \Delta _4>\Delta $, then by interchanging the role of
the pairs $(13)$ and $(24)$, one may always assume that $\Delta _1 + \Delta _3
> \Delta$. Thus, only the case where both $\Delta _1 + \Delta _3 \leq \Delta$
{\it and} $\Delta _2 + \Delta _4 \leq \Delta$ will not lead to a truncated
$k$-series.

The structure of quantum field theory on $\AdS$ space by itself does not
guarantee a condition like (4.1). However, quantum field theory on $\AdS\times S$
will be restricted by the tensor product rules of the finite-dimensional
representations of the sphere isometry group, and this produces inequalities
such as (4.1). Notice that the case of gauge boson exchange falls into this
category [11].

\medskip

\noindent
{\bf c. Integer $\delta$ : Amplitudes from a Universal Function}

\smallskip

The amplitude $B_R(x,y)$ of (3.24) may be expressed in terms of $\rho (\tau)$
and a universal function that will not depend upon any of the
dimensions of the problem. To see this, we introduce the notation 
$$
X=x^2, \qquad Y=y^2, \qquad Z=\half (x-y)^2
\eqno (4.2)
$$
and temporarily declare the variables $X$, $Y$ and $Z$ independent, setting
them to the values of (4.2) only later. Defining also the combinations
$$
\eqalign{
n_x & = \delta - \Delta _1 - \Delta _2 
      = \12 (- \Delta _1 - \Delta _2 +\Delta _3 +\Delta _4)
\cr
n_y & = \delta - \Delta _1 - \Delta _4 
      = \12 (- \Delta _1 + \Delta _2 +\Delta _3 -\Delta _4)
\cr
n_z & = \delta - \Delta _3 -1 = \delta -\Delta _1 -n_x-n_y-1\, ,
\cr}
\eqno (4.3)
$$
we have
$$
B_R (x,y) = {(-)^{\delta + \Delta _1+1} \over 2^{n_z} \Gamma (\delta -\Delta _1)}\ 
{\partial ^{n_x + n_y + n_z} \hat B_R (X,Y,Z)
             \over \partial X ^{n_x} \partial Y ^{n_y} \partial Z ^{n_z}}
\bigg | _{X=x^2,\ Y=y^2, \ Z=\half (x-y)^2}
\eqno (4.4)
$$
where the function $\hat B_R(X,Y,Z)$ is defined by
$$
\hat B_R (X,Y,Z) = \int _0 ^\infty {du \over u} \int _0 ^\infty {dv \over v}\ 
{\rho \big ({uv \over u + v +uv}\big ) \over uX + vY +2Z}\, .
\eqno (4.5)
$$

\medskip

For expression (4.4) to make sense, the number of derivatives with
respect to each variable $X$, $Y$ and $Z$ must be positive or zero, so that we
must have
$n_x\geq 0$, $n_y \geq 0$ and $n_z \geq 0$. From the convergence condition
(3.2b), it is clear that $n_z \geq 0$. The conditions $n_x\geq 0$ and $n_y \geq
0$ require that $\Delta _3 - \Delta _1 \geq |\Delta _4 - \Delta _2|$, which may
always be achieved simply by interchanging the roles of the external legs of the
amplitude.

\medskip

There is a special case (which does not occur in the $AdS_5 \times S_5$ theory)
where the assumptions of \S~b above may conflict with the assumptions of the
preceding paragraph. This is when $\Delta _1 + \Delta _3 > \Delta \geq \Delta
_2 + \Delta _4$. The roles of the pairs $(13)$ and $(24)$ may have to be
interchanged in order to obtain the inequalities in this ordering, and it may
not be possible then to choose  $\Delta _3 - \Delta _1 \geq |\Delta _4 -
\Delta _2|$ as well. If needed, this case is best handled separately with
methods parallel to the ones to be used below.

\medskip

To calculate $B_R$, it suffices to obtain $\hat B_R$, a task on which we now
concentrate. Performing a change of variables $\tau = uv/(u+v+uv)$ and $\lambda =
(v-u)/(v+u)$, one finds that
$\hat B_R$ may be expressed in terms of a single integral over $\tau$ of
$\rho(\tau)$ and a universal function $I(\tau; s,t)$, as follows
$$
\hat B_R (X,Y,Z) = {1 \over X+Y} \int _0 ^1 {d \tau \over \tau} \rho (\tau ) \ 
I\big (\tau;{Z\over X+Y}, {X-Y \over X+Y}\big )\, . 
\eqno (4.6) 
$$
The universal function is elementary and given by
$$
\eqalignno{
I(\tau; s,t) & = \int _{-1} ^{+1} d \lambda \ {1 \over \tau (1-\lambda
t) + s(1-\lambda ^2) (1-\tau)}
& (4.7a) \cr
& =
 {1\over \sqrt {\omega ^2 -(1-t^2)\tau^2}}
\biggl \{ -\ln (1-t^2)\tau ^2 + 2 \ln (\omega +\sqrt{\omega ^2 -(1-t^2)\tau^2})
\biggr
\}\, , & (4.7b)\cr
}
$$
where we have set $\omega = 2s(1-\tau) +\tau$ for short.
This form of the amplitude will be very useful when deriving convergent series
expansions and asymptotics.

\bigskip
\bigbreak

\centerline{\bf 5. SHORT DISTANCE EXPANSIONS}

\bigskip

The main purpose of this section is to study the short distance limits of
scalar exchange correlation in the direct channel $x_2 - x_4 \to 0$, and the
crossed channel $x_2-x_3 \to 0$. We are interested in the sense in which the
amplitude admits double OPE expansions of the type (1.1) in these limits. This
question will be analyzed using the series expansion of $\rho (\tau)$ of (3.24)
and then estimating the contribution of each term to the amplitude $B_R(x,y)$
via (3.23). However, we will first discuss the leading term in the direct
channel using the simple method discussed in  [7] which works  from the
original expression (3.1b) for the amplitude, before any processing. This
discussion will bring in some physical aspects of the AdS/CFT correspondence,
and it will provide checks on the intricate manipulations used to derive (3.23)
and (3.24) and then short distance expansions.

\bigskip

\noindent
{\bf a. Leading term in direct channel $x_{24} \to 0$.}

\medskip

The intuitive idea behind the simple method of [7] is that the region of
$AdS_{d+1}$ which gives the dominant contribution in the direct channel limit
is the region where $\vec{z}$ is near $\vec{x_1}$ and $\vec{w}$ near $\vec
{x_2}$. Thus, we consider an expansion of the bulk-to-bulk scalar propagator in
(3.1b) about these points. The leading term for $|\vec{x_{12}}| \gg
|\vec{x_{13}}|,\ |\vec{x_{24}}|$ is obtained from the term in the series 
expansion of the hypergeometric function with slowest fall-off as $u\to
\infty$, so that our approximation is
$$
\xi ^\Delta F({\Delta \over 2}, {\Delta +1 \over 2}; \Delta - \d +1; \xi ^2)
\ \sim \ \xi ^\Delta \ \sim \ 
{(2z_0 w_0)^\Delta \over | \vec{x_{12}}|^{2 \Delta }}\, .
\eqno (5.1)
$$
We insert this in (3.1b) and obtain the factorized form 
$$
A(x_i) \sim {2^\Delta \over |\vec{x_{12}}|^{2 \Delta}}
  V(\Delta _1, \Delta _3, \Delta , |\vec{x_{13}}|) 
  V(\Delta _2, \Delta _4, \Delta , |\vec{x_{24}}|)\, , 
\eqno (5.2a)
$$
where
$$
\eqalign{
V(\Delta _i, \Delta _j, \Delta , |\vec{x_{ij}}|)
& = \int {d^{d+1} z\over z_0 ^{d+1}} {z_0 ^{\Delta _i + \Delta _j + \Delta}
\over (z-x_i)^{2 \Delta _i} (z-x_j) ^{2 \Delta _j}} 
\cr
& = {T \ \Gamma (\12(\Delta + \Delta _i +\Delta _j) -\d) 
    \over
    \Gamma (\Delta _i) \Gamma (\Delta _j) \Gamma (\Delta)
   |\vec{x_{ij}}|^{\Delta _i + \Delta _j - \Delta}}
\cr}
\eqno (5.2b)
$$
with 
$$
T=\Gamma \big (\12 (\Delta_i + \Delta_j - \Delta) \big )
     \Gamma \big (\12 (\Delta + \Delta_i  - \Delta_j) \big )
     \Gamma \big (\12 (\Delta  + \Delta_j - \Delta_i ) \big )\, .
\eqno (5.2c)
$$
It is important that $V(\Delta _i, \Delta _j, \Delta, |x_{ij}|)$ differs from
the 3-point correlator of the CFT operators $\O_{\Delta _i}$, $\O_{\Delta _j}$
and $\O_{\Delta }$ only by normalization factors, including those factored out
in (3.1a), and that the result above for the integral in (5.2) was obtained
directly from (22-23) of [20], where the 3-point correlators were
systematically studied. In fact, when normalization factors in (3.1a) are
included, one can show that the leading term in the 4-point correlator is {\it
exactly} the result expected for the contribution of the primary operator
$\O_{\Delta}$ in the double OPE of (1.1). This confirms that the AdS/CFT
correspondence satisfies  expected non-linear properties in the boundary
theory.

\medskip

We observe that poles in $\Gamma$-functions in (5.2) correspond to divergences
in the AdS integral for the correlator $\langle \O _{\Delta _i} \O_{\Delta _j}
\O _\Delta \rangle$, and in particular this correlator diverges unless 
$|\Delta _i - \Delta _j| < \Delta < \Delta _i + \Delta _j$. We have already
observed (see (3.2a)) that the 4-point amplitude also diverges when the lower
limit is not obeyed, and we now see that the upper limit is the triangle
inequality (4.1), which is the condition for truncation of the infinite series
in the spectral density. The 4-point correlator converges if this inequality
is violated, but we cannot expect that the direct channel limit conforms to a
double OPE when the 3-point function diverges. Thus, we should expect a
qualitatively different asymptotic behavior, and this will be confirmed by our
systematic analysis below.

\bigskip

\noindent
{\bf b. Expansion in the Direct Channel for $\Delta < \Delta _1+ \Delta _3$.}

\medskip

Here, we shall concentrate on the direct channel OPE, in which $x-y \to 0$, while
keeping $x$ and $y$ finite. In this regime, $X-Y$ and $Z$ both tend to 0, so
that $s \to 0$ and $t\to 0$ in (4.6b). We begin by carrying out a convergent
expansion in powers of $(X-Y)$, i.e. in terms of $t$, which is valid for both
$\Delta < \Delta _1 + \Delta _3$ (this subsection) and $\Delta \geq \Delta _1
+\Delta _3$ (subsection c.).

\medskip

\noindent
{\it Expansion in powers of $t$}

\medskip

The universal function $I(\tau;s,t)$ of (4.6b) admits a simple convergent series
expansion in powers of $t$, given as follows
$$
\eqalignno{
I(\tau;s,t) & =  \sum _{m=0} ^\infty  t^{2m} I_{2m} (\tau ; s)
  & (5.3a) \cr
I_{2m} (\tau ; s) & = 2\int _0 ^1 d \lambda 
  {\lambda ^{2m} \tau ^{2m} \over [\tau + s (1-\tau) (1-\lambda ^2)]^{2m+1}}\, .
  & (5.3b) \cr}
$$
Since $I_{2m}(\tau;s) \leq I_0 (\tau;s)$ for all $\tau \in [0,1]$ and $s\geq
0$, the radius of convergence in $t$ of (5.3a) is larger or equal to 1.

As a result, the function $\hat B_R(X,Y,Z)$ (and thus $B_R(x,y)$) admits an
expansion in powers of $t$ as well, and we have
$$
\eqalignno{
\hat B_R (X,Y,Z) 
  & = {1 \over X+Y} \sum _{m=0} ^\infty t^{2m} B^{(2m)} ({Z \over X+Y}) 
  & (5.4a) \cr
B^{(2m)} (s) & = \int _0 ^1 {d\tau \over \tau} \ \rho (\tau) \ I_{2m}(\tau;s)\, .
  & (5.4b) \cr}
$$
It remains to compute the partial amplitudes $B^{(2m)}(s)$.

\medskip

\noindent
{\it The Exact form of $\rho (\tau)$.}

\medskip

In order to analyze the $s$-dependence of the partial amplitudes $B_{2m}(s)$, we
need good control over the $\tau$-dependence of $\rho (\tau)$. We consider first
the case where the $k$-sum of $\rho_0(\tau)$ truncates to a finite sum, which
occurs when $c-a \leq 0$, or equivalently $\Delta <\Delta _1 + \Delta _3$. 
In this case, the infinite series of (3.29) collapses to a finite sum. Also,
precisely for these values does the hypergeometric function in (3.30) become a
rational function of $\tau$, and $\rho _\ell ^{(k)}(\tau)$ a polynomial in
$\tau$ of degree $\Delta _1 -1$. Both are given by 
$$
\eqalignno{
\rho _\ell (\tau) 
& = 
\sum _{k=0} ^{a-c} 
  {\Gamma (1+a-c) \Gamma (c-b+k) \Gamma (c) 
     \over
   \Gamma (1+a-c-k) \Gamma (c-b) \Gamma (c+k) \Gamma (k+1)} \ (-)^k
\ \rho _\ell ^{(k)} (\tau)
& (5.5a) \cr
\rho _\ell ^{(k)} (\tau)
& = 
 \sum _{n=0} ^{a-c-k} 
{\Gamma (a-c+1 -k) \Gamma (\Delta -b+k+n) \Gamma (1+a-c+\ell-n-k)
\over \Gamma (\Delta _1+\ell )  \Gamma (1+a-c-n-k) \Gamma (n+1)}
\tau ^{\Delta -b +k+n}\qquad \qquad & (5.5b) \cr}
$$
where the latter may be established with the help of [23]. 
Finally, putting together (3.26) with (5.5), we obtain a complete expression
for $\rho(\tau)$, as follows,
$$
\eqalignno{
\rho (\tau) 
& = 
\sum _{k=0} ^{a-c} \ \sum _{n=0} ^{a-c-k} 
  {(-)^k E_{k+n} \Gamma (c-b+k) \Gamma (\Delta -b+k+n) 
     \over
    \Gamma (c+k) \Gamma (c-b) \Gamma (k+1)\Gamma (n+1) } 
\tau ^{\Delta -b+k+n}
& (5.6a) \cr
E_k
& = 
   {\Gamma (\delta -\d-1-a+c+k) \Gamma (\Delta _3) \Gamma (a-c+1 )   
    \Gamma (c) 
\over 
    \Gamma (\Delta _1 ) \Gamma (\delta -\d) 
    \Gamma (b+k) }
 & (5.6b) \cr}
$$
Polynomial behavior of $\rho (\tau)$ is ideal for the evaluation of the 
asymptotics of the amplitudes, as we shall see next.

\medskip

\noindent
{\it The Calculation of $B^{(2m)}(s)$.}

\medskip

For $\Delta < \Delta _1 + \Delta _3$, we have $c-a \leq 0$, and thus $\rho
_\ell(\tau)$ and $\rho(\tau)$ are polynomial in $\tau$, given by (5.5-6). Notice
also from these expressions that $\rho _\ell ^{(k)}(\tau)$ and thus $\rho (\tau)$
vanishes at $\tau=0$, so that the integral in (5.4b) is always convergent. It
remains to evaluate, for $p\geq 1$, the integrals
$$
B^{(2m)} _p (s) \equiv \int _0 ^1 d\tau \ \tau ^{p-1} I_{2m}(\tau;s)\, .
\eqno (5.7)
$$
Using (5.3b) and interchanging the orders of integration of $\tau$ and
$\lambda$, we find
$$
B^{(2m)}_p (s) 
 = 2 \int _0 ^1 d \lambda \ \lambda ^{2m} \int _0 ^1 d\tau \ \tau ^{2m+ p-1}
\bigl [ \tau + s (1-\tau) (1-\lambda ^2) \bigr ] ^{-1-2m}\, .
\eqno (5.8)
$$
The $\tau$-integral is readily recognized as a hypergeometric function,
$$
B^{(2m)}_p (s) 
 = {2 \over 2m+p} \int _0 ^1 d \lambda \ \lambda ^{2m}                   
F\bigl (1,2m+1;2m+p+1;1 - s(1-\lambda ^2) \bigr )\, .
\eqno (5.9)
$$
Using [24], we see that this (somewhat complicated) elementary function $F$ has
a simple convergent expansion for small $s$. We recall from [24] that
$$
\eqalign{
F(1,& 2m+1;2m+p+1;1-z) \cr 
& = {2m+p \over \Gamma (p) \Gamma (2m+1)}
\biggl \{\sum _{n=0} ^{p-2} (-)^n \Gamma (2m+n+1) \Gamma (p-n-1) z^n
\cr
& \qquad +(-)^{p-1} \sum _{n=0} ^\infty {\Gamma (2m+p+n) \over \Gamma (n+1)}
\bigl (\psi (n+1) - \psi (2m+p+n) -\ln z\bigr ) z^{n+p-1} \biggr \}\, .
\cr}
\eqno (5.10)
$$
To compute $B^{(2m)}_p(s)$, we set $z=s(1-\lambda ^2)$ and perform the
$\lambda$-integrals explicitly, with the help of 
$$
\eqalign{
\int _0 ^1 d \lambda \lambda ^{2m} (1-\lambda ^2)^n &= 
{\Gamma (m+\half) \Gamma (n+1) \over 2\Gamma (m+n+{3\over 2})}
 \cr
\int _0 ^1 d \lambda \lambda ^{2m} (1-\lambda ^2)^n \ln (1-\lambda^2) &= 
{\Gamma (m+\half) \Gamma (n+1) \over 2\Gamma (m+n+{3\over 2})}\bigl (\psi
(n+1)-\psi(m+n+{3\over2})\bigr )
 \cr}
\eqno (5.11)
$$
Putting all together, we have
$$
\eqalign{
B^{(2m)}_p 
& = {\Gamma (m+\half) \over \Gamma (p) \Gamma (2m+1)}
\biggl \{\sum _{n=0} ^{p-2} (-)^n 
  {\Gamma (2m+n+1) \Gamma (p-n-1) \Gamma (n+1) \over \Gamma (m+n+{3\over 2})}
s^n
\cr
& \qquad \qquad +(-)^{p-1} \sum _{n=0} ^\infty 
{\Gamma (2m+p+n) \Gamma(n+p) \over \Gamma (m+n+p+\half)\Gamma (n+1)}
\bigl (h_n -\ln s\bigr ) s^{n+p-1}  \biggr \}\,.
\cr}
\eqno (5.12)
$$
where the combinations $h_n$ are defined by
$$
h_n = \psi (n+1) - \psi (n+p) +\psi (m+n+p +\half) - \psi (2m+p+n)\, .
\eqno (5.13)
$$
The coefficient function that multiplies the $\ln s$ contribution is manifestly
a hypergeometric function, given by
$$
B^{(2m)} _p (s) \bigg |_{ \ln s}
= (-)^p s^{p-1} \ln s \ 
  {\Gamma (m+\half) \Gamma (2m+p)\over \Gamma (2m+1) \Gamma (m+p+\half)}
F(2m+p,p;m+p+\half;s)
\eqno (5.14)
$$
Combining (5.4), (5.6) and (5.12), we finally obtain a complete expression for
$B^{(2m)}(s)$ in terms of $s$, by
$$
B^{(2m)}(s) 
 = 
\sum _{k=0} ^{a-c} \ \sum _{n=0} ^{a-c-k} 
  {(-)^k E_{k+n} \Gamma (c-b+k) \Gamma (\Delta -b+k+n) 
     \over
    \Gamma (c+k) \Gamma (c-b) \Gamma (k+1)\Gamma (n+1) } 
B_{\Delta -b+k+n} ^{(2m)}(s)\, .
\eqno (5.15)
$$
The complete logarithmic contribution may easily be read off from (5.14) and
(5.15), and we find
$$
\eqalign{
B^{(2m)} (s) \bigg |_{ \ln s}
& = \ln s \
\sum _{k=0} ^{a-c} \ \sum _{n=0} ^{a-c-k} 
  { E_{k+n} \Gamma (c-b+k) \Gamma (\Delta -b+k+n) 
     \over
    \Gamma (c+k) \Gamma (c-b) \Gamma (k+1)\Gamma (n+1) } 
\cr & \qquad 
(-)^{\Delta -b+n} s^{\Delta -b+k+n-1}  
  {\Gamma (m+\half) \Gamma (2m+\Delta -b+k+n)\over \Gamma (2m+1) \Gamma
(m+\Delta -b+k+n+\half)}
\cr & \qquad \qquad 
 F(2m+\Delta -b+k+n,\Delta -b+k+n;m+\Delta -b+k+n+\half;s)\, .
\cr}
\eqno (5.16)
$$
The contribution of $B^{(2m}(s)$ to the full amplitude is then gotten by
combining (5.15) and (5.13) with (5.4a) and (4.4), in particular, all the
logarithmic contributions arise from (5.15).

\medskip

\noindent
{\bf c. Matching of Direct Channel Results with OPE Predictions}

\medskip

It is worthwhile to calculate the leading term of $A(x_i)$ directly from
the formulas above as a check on the long development of the previous
sections. The process is tedious so we shall present only an
outline of the steps to be followed.

(i) The relation [11] between the variables $s$, $t$ used above and
conformal
invariant functions of the original coordinates $x_i$ is
$$
s=\half {x_{13}^2 x_{24}^2 \over x_{12}^2 x_{34}^2 + x_{14}^2 x_{23}^2}
\qquad \qquad
t= {x_{12}^2 x_{34}^2 - x_{14}^2 x_{23}^2 \over x_{12}^2 x_{34}^2 +
x_{14}^2
x_{23}^2}
\eqno (5.17)
$$
In the direct channel limit, $x_{13}$ and $x_{24}$ are small, and all
other large $x_{ij}^2$ are equal to $x_{12}^2$ to leading order. The leading
behaviors of $s$ and $t$ are
$$
s\sim {x_{13}^2\ x_{24}^2 \over 4 x_{12}^4}
\qquad \qquad
t \sim {(x_{31})_\mu J_{\mu \nu} (x_{12}) (x_{42})_\nu \over x_{12}^2}
\eqno (5.18)
$$
where $J_{\mu \nu} (y) = \delta _{\mu \nu} - 2 y_\mu y_\nu /y^2$ is the
well-known inversion Jacobian. We see that $s$ and $t$ both vanish in the
limit, but $t$ vanishes more slowly than $s$. For the leading term, it is thus
sufficient to consider only the $m=0$ term in (5.1a) and only $B^{(0)}(s)$
and $B^{(0)}_p(s) $ in (5.12) and (5.16).

(ii) To simplify somewhat we consider only the special case where
$\Delta_1=\Delta_3$ and $\Delta_2=\Delta_4$, and we further assume that $\Delta$
is an even integer, as implied by the selection rules of the $\AdS_5\times S_5$
theory discussed in Sec 1. It is then the case that the minimum value of the
subscript $p$ in  $B_p^{(2m)}$ in (5.12) and (5.15) is $p_0=\Delta/2$, and the
entire singular contribution as $x_{24}\to 0$ is given by the derivatives of
(4.4) applied to $B^{(2m)}(s) \big |_{\ ln  s}$ in (5.16). For the leading
singularity we can further restrict to 
$B^{(0)}(s)|_{\ln  s}$, as shown in (i) above, and we can keep only the $k=n=0$
contribution in (5.16) with the hypergeometric function replaced by $F \to 1$.
The higher powers of $s$ we have dropped give only non-leading terms in the
direct-channel limit.

(iii) Turning to (4.3-4.4) we see that $n_x=n_y=0$ and $n_z=\Delta_2-1$
with our assumptions. We must then compute the $Z$-derivatives in (4.4),
whose leading singularity at small $s$ is given by 
$$
{\partial ^{n_z} \over \partial z^{n_z}} \bigl (\ln s \ s^{p_0-1}\bigr )=
{(-)^{\Delta_2-\half\Delta+1} 
\Gamma(\half \Delta) \Gamma(\Delta_2-\half\Delta)
\over
(X+Y)^{\Delta_2-1}}
{1 \over s^{\Delta_2 - \half\Delta}}
 \eqno (5.19)
$$
One can see that this has the same singularity as (5.2) as $x_{24} \to 0$. Note
that
$$
X+Y=(x_4'-x_3')^2 + (x_2'-x_3')^2= {x_{12}^2 x_{34}^2 +x_{23}^2 x_{14}^2 \over
x_{13}^2 x_{12}^2 x_{14}^2}
\eqno (5.20)
$$
The coordinate dependence of the leading term is obtained by combining the
factors in (5.19) with those from inversion in (3.3). The result is
$$
{1 \over |x_{12}|^{2\Delta} |x_{13}|^{2\Delta_1-\Delta}
|x_{24}|^{2\Delta_2-1}}
\eqno (5.21)
$$
in agreement with (5.2). To compute its coefficient one must combine all
$\Gamma$-functions and powers of $2$ and $(-)$ from (5.16), (5.5b), (5.2a),
(4.4), and (3.23a). The result agrees precisely with (5.2) and the double 
OPE.

\medskip

To close this subsection we note that not only the leading term, but also
the full asymptotic series of the direct channel limit of the correlation
function is contained in the formulas (5.2) and (5.16). In particular the most
singular logarithmic term is generically
$$
A(x_i) \sim {1 \over |x_{13}|^{2(\Delta_1-\Delta_2)} |x_{12}|^{4\Delta_2} } 
            \ln {x_{13}^2 x_{24}^2\over x_{12}^4}.
\eqno (5.22)
$$
Note that the leading logarithmic singularity always arises multiplied by a
regular power term in the expansion.

\bigskip

\noindent
{\bf d. Expansion in the Direct Channel $x_{24} \to 0$ for $\Delta \geq
\Delta _1 + \Delta _3$.}

\medskip

Recall that the expansions in powers of $t$ in (5.3) and (5.4) continue to hold
in this case.

\medskip

In this case, we have $c-a\geq 1$ and the series in (3.29) for $\rho (\tau)$
truly has an infinite number of terms. The coefficients $Q_p$ of (3.32) in the
series expressions for $\rho _\ell$ of (3.31) behave as $Q_p \sim 1/p^{c-a}$ as
$p\to\infty$, so that the series of (3.31) will be uniformly convergent
throughout $\tau \in [0,1]$ for $c-a \geq 2$, and uniformly convergent
throughout $\tau \in [0,1-\epsilon]$, $\epsilon >0$ for $c-a=1$. Clearly, the
series defines a function that is analytic around $\tau =1$ by (3.31). 
At $\tau =0$, $\tau$-derivatives of $\rho _\ell$ up to order $c-a+\Delta _1-1$
will be   finite, but an extra derivative leads to a divergence, signaling
non-analyticity at $\tau=0$. Thus, the analytic continuation of the sum (3.31)
and its asymptotics around $\tau =0$ have to be extracted with care.

\medskip

We use standard methods of analytic continuation and asymptotics, applied to
string loop amplitudes in [25], to obtain the desired asymptotics. For
simplicity, we restrict attention to $\rho _0(\tau)$, the generalization to
other values of $\ell$ being available from (3.31). 

\medskip

Non-analyticity at $\tau =0$ arises from the slow rate of convergence of the
$p$-series. From (3.32a), it is manifest that the large $p$ behavior of the
contribution of a single term of order $k$ in the series (3.32a) is given by
$1/p^{c-a+k}$. Thus, the  worst non-analyticities in $\rho _0(\tau)$ arise from
the terms in the series (3.32a) with the lowest values of $k$. Assuming that we
are interested in the asymptotics of $\rho (\tau)$ up to a given order, say
$K$, we may isolate the first $K$ terms in the series (3.32a), treating these
exactly, and then bound the remaining infinite series, which now has improved
analytic behavior at $\tau=0$. Thus, we define
$$
Q_p = Q_p ^K + 
  \sum _{k=0} ^{K-1} 
   {\Gamma (c-a+k) \Gamma (c-b+k) \Gamma (c) \Gamma (\Delta -b+k)         
    \Gamma (p+\Delta _1) 
    \over 
    \Gamma (c-a) \Gamma (c-b) \Gamma (c+k) \Gamma (k+1) 
    \Gamma (\Delta -b+k+p+1) \Gamma (\Delta _1)}
\eqno (5.23)
$$
where the large $p$ behavior of $Q_p ^K$ is now given by $Q_p \sim 1/p^{c-a+K}$.
As a result, the contribution to $\rho_0 (\tau)$ from $Q_p^K$ will have improved
analytic behavior at $\tau=0$ and will be differentiable $c-a+\Delta _1 +K-1$ 
times.

\medskip

The leading asymptotics of $\rho _0(\tau)$ may now be worked out by obtaining
that of the leading $K$ terms of (5.23). (Equivalently, one can  proceed to
obtain the asymptotics from (3.30), and making use of (5.10), but we find it
instructive to follow the route below.) For each $k$ with
$0\leq k\leq K-1$, we perform the $p$-sum exactly. Thus, we need to resum the
series
$$
\sum _{p=0} ^\infty { \Gamma (p+\Delta _1) \over \Gamma (\Delta -b+k+p+1)}\ 
(1-\tau)^p\, ,
\eqno (5.24)
$$
which is most easily carried out by decomposing the ratio of $\Gamma$-functions
in terms of its simple poles, as follows
$$
{ \Gamma (p+\Delta _1) \over \Gamma (\Delta -b+k+p+1)}
=
\sum _{q=0} ^{\Delta -b-\Delta _1 +k}
{(-)^q \over
  \Gamma (\Delta -b -\Delta _1 +k -q+1) \ q!}
\ {1 \over \Delta _1 +q +p}\, .
\eqno (5.25)
$$
Making use of the following $p$-sums
$$
\sum _{p=0} ^\infty {(1-\tau)^p \over \Delta _1 + q +p}
= (1-\tau)^{-\Delta _1 -q} \biggl (
-\ln \tau - \sum _{p=1} ^{\Delta _1 +q-1} {1 \over p} (1-\tau)^p \biggr )\, ,
\eqno (5.26)
$$
we may readily extract the logarithmic behavior of (5.24) by resumming in (5.25)
only the logarithmic contributions of (5.26).
Thus, putting all together, we have
$$
\eqalign{
\sum _{p=0} ^\infty { \Gamma (p+\Delta _1) \over \Gamma (\Delta -b+k+p+1)}\ 
(1-\tau)^p
& = {(-\tau)^{\Delta -b-\Delta _1 +k} (1-\tau)^{-\Delta +b -k}
\over 
\Gamma (\Delta -b - \Delta _1 +k+1)}
\ \ln \tau + {\rm analytic} \cr
& \sim {(-\tau)^{\Delta -b-\Delta _1+k} 
\over \Gamma (\Delta -b - \Delta _1 +k+1)} \ \ln \tau
\cr}
\eqno (5.27)
$$
Despite its appearance, expression (5.26) is perfectly regular
and analytic in $\tau$ around $\tau=1$, since the original sums were. Thus, the
novel feature of $\rho (\tau)$ when $\Delta \geq \Delta _1 +\Delta _3$ is the
appearance of logarithmic singularities around $\tau =0$. Also, from the
arguments above, we see that no other non-analyticities take place for $\tau
\in [0,1]$.

\medskip

\noindent
{\it Calculation of $B^{(2m)}(s)$}

\medskip

The novel feature in dealing with the case $\Delta \geq \Delta _1 + 
\Delta _3$ is the appearance of the integrals (for $p\geq 1$) of the type
$$
\tilde B _p ^{(2m)} (s) \equiv
  \int _0 ^1 d\tau \ \tau ^{p-1} \ \ln \tau \ I_{2m} (\tau;s)\, ,
\eqno (5.28)
$$
where $p$ takes on the values of $\Delta -b-\Delta _1 +k$ of (5.27).
They may be evaluated by starting from the function $B_p ^{(2m)}(s)$ for 
arbitrary real values of $p$ and then taking the derivative with respect
to $p$ :
$$
\tilde B_p ^{(2m)} (s) = {\partial \over \partial p} B_p ^{(2m)}(s).
\eqno (5.29)
$$ 
Eqs. (5.8) and (5.9) still hold when $p$ is real instead of integer, but an
analytic continuation that is more general than (5.10) is now needed. This is
given by [26], and may be re-expressed by expanding the hypergeometric
functions in a power series, as follows
$$
\eqalign{
{1 \over 2m+p} & F(1, 2m+1;2m+p+1;1-z) \cr 
& = {\Gamma (1-p) \over \Gamma (2m+1)}
\biggl \{- \sum _{n=0} ^\infty  {\Gamma (2m+n+1) \over \Gamma (2-p+n)} z^n
 + \sum _{n=0} ^\infty {\Gamma (2m+p+n) \over \Gamma (n+1)}
 z^{n+p-1} \biggr \}\, .
\cr}
\eqno (5.30)
$$
To compute $B_p ^{(2m)}(s)$ for arbitrary real $p$, we set $z=s(1-\lambda ^2)$
and perform the $\lambda$-integrals explicitly, using (5.11), 
$$
\eqalign{
B^{(2m)}_p 
 = {\Gamma (m+\half) \Gamma (1-p)\over  \Gamma (2m+1)}
\biggl \{& -\sum _{n=0} ^\infty  
  {\Gamma (2m+n+1) \Gamma (n+1) \over \Gamma (2-p+n) \Gamma (m+n+{3\over 2})}
s^n
\cr
& + \sum _{n=0} ^\infty 
{\Gamma (2m+p+n) \Gamma(n+p) \over \Gamma (m+n+p+\half)\Gamma (n+1)}
s^{n+p-1}  \biggr \}\,.
\cr}
\eqno (5.31)
$$
The prefactor $\Gamma (1-p)$ will produce a pole at positive integer values of
$p$, but the combined expression is regular in view of the fact that the two
series inside of the brace of (5.31) cancel one another when $p$ is a positive
integer. It is straightforward to obtain the derivative with respect to $p$ of
this expression, and then to set $p$ to a positive integer. The result is as
follows
$$
\eqalign{
\tilde B^{(2m)}_p 
& = {\Gamma (m+\half) \over \Gamma (p) \Gamma (2m+1)}
\biggl \{\sum _{n=0} ^{p-2} (-s)^n 
  {\Gamma (2m+n+1) \Gamma (p-n-1) \Gamma (n+1) \over \Gamma (m+n+{3\over 2})}
\bigl ( \psi (p-n-1) - \psi (p) \bigr )
\cr
& \qquad \qquad +(-)^p \sum _{n=0} ^\infty 
{\Gamma (2m+p+n) \Gamma(n+p) \over \Gamma (m+n+p+\half)\Gamma (n+1)}
\ H_n (s) \ s^{n+p-1}  \biggr \}\,.
\cr}
\eqno (5.32)
$$
where the functions $H_n(s)$ are defined by
$$
\eqalign{
H_n (s) = \half & (\ln s)^2 + \big [ h_n + \psi (n+1) - \psi (p) \big ] \ln s
\cr &
+ \half h_n' 
+ \half h_n \big [ h_n + 2\psi (n+1) - 2 \psi (p) \big ]
\cr}
\eqno (5.33)
$$
where  $h_n$ was defined in (5.13) and $h_n'$ is given by
$$
h_n' = \psi '(n+1) - \psi '(n+p) +\psi '(m+n+p +\half) - \psi '(2m+p+n)\, .
\eqno (5.34)
$$
In particular the logarithm square term may be isolated and yields
$$
\tilde B^{(2m)} _p (s) \bigg |_{ (\ln s)^2}
= (-)^p s^{p-1} (\ln s)^2  \ 
  {\Gamma (m+\half) \Gamma (2m+p)\over \Gamma (2m+1) \Gamma (m+p+\half)}
F(2m+p,p;m+p+\half;s)\, .
\eqno (5.35)
$$
We see that the presence of logarithmic singularities in $\rho (\tau)$
generates logarithm squared terms in the amplitude when $\Delta _1 + \Delta _3
\geq \Delta$.

\bigskip

\noindent
{\bf e. Expansions in the Crossed Channel $x_3-x_2 \to0$}

\medskip

In the crossed channel, the expansion is given in terms of $y\to 0$, while $x$
is kept fixed, corresponding to $s\to \12$ and $t \to +1$. We shall show here
that the amplitude behaves in a universal way in this limit, as long as the
dimensions $\Delta _i$ and $\Delta$, as well as $\delta$ are integers. In
particular, there is no need here to make assumptions about the relative
magnitudes of the dimensions, as long as they obey the convergence requirements
of (3.2). 

\medskip

Our starting point is the universal formula (4.7). Around $s=\12$ and $t=1$, 
the composite variable $\omega$ behaves as $\omega \sim 1$.
As a result, the universal function $I(\tau;s,t)$ may be split into a
logarithmically singular and a regular analytic part, $I=I_{{\rm sing}}+I_{{\rm
reg}}$ as follows,
$$
\eqalignno{
I_{{\rm sing}} (\tau;s,t) 
& = {- \ln (1-t^2) \over \sqrt {\omega ^2 - (1-t^2)\tau^2}}
& (5.36a) \cr
&&\cr
I_{{\rm reg}} (\tau;s,t)
& = {- \ln \tau ^2 +2 \ln \{\omega +\sqrt {\omega ^2 - (1-t^2)\tau^2}\} 
\over
\sqrt {\omega ^2 - (1-t^2)\tau^2}} & (5.36b) \cr}
$$
where $\omega = 1+(2s-1)(1-\tau)$ for short, as in (4.7).
The square root in $I_{{\rm sing}}$ as well as the full function $I_{{\rm
reg}}$ admits convergent series expansions in powers of $(1-t^2)$ and $(2s-1)$,
with coefficients which are regular functions of $\tau$, leading to convergent
$\tau$-integrations versus the spectral function $\rho(\tau)$. As a result,
$I_{{\rm reg}}$ contributes to the amplitude $B_R(x,y)$ through a convergent
power series in $(1-t^2)$ and $(2s-1)$. 
$I_{{\rm sing}}$ on the other hand, has
an overall factor of $\ln (1-t^2)$, and multiplies a function that also admits a
convergent power series in $(1-t^2)$ and $(2s-1)$. 
The $\ln (1-t^2)$ factor simply gives the factor
$\ln Y/X$ in  $\hat B_R(X,Y,Z)$ and higher powers of $(1-t^2)$ can be dropped if we
are interested only in the leading term as $Y\to 0$. The power series in $(2s-1)$ is
then a geometric series, and we obtain
$$
\hat B_R(X,Y,Z) \approx {-1\over X} \ln {Y\over X} \sum_{n=0}^\infty (1-2s)^n \int_0^1
{d\tau\over \tau}\rho(\tau) (1-\tau)^n + {1\over X} {\cal O}(Y/X) .
\eqno(5.37)
$$
This must be inserted in (4.4) and differentiated, and the result depends on whether
$n_y$ is positive or zero. If $n_y=0$ the leading singularity in $Y$ is logarithmic, and
its coefficient comes from the $n=n_z$ term of the series in (5.37). We give the final
result for the leading term in the case $\Delta_1=\Delta_3$ and $\Delta_2=\Delta_4$,
which is 
$$
A(x_c) \sim {1\over x_{31}{}^{2\Delta_1}} {1\over x_{24}{}^{2\Delta_2}} \ln
\Bigl({x_{23} x_{41}\over x_{21} x_{34}}\Bigr) \int_0^1
{d\tau\over \tau}\rho(\tau) (1-\tau)^{\Delta_2-1}. \eqno(5.38)
$$
When $n_y >0$, the leading term contains the power singularity $1/(x_{23})^{2n_y}$. 

\bigbreak

\centerline{{\bf 6. CONCLUSIONS AND OUTLOOK}}

\bigskip

In this paper we have studied the contribution of the exchange of a bulk
scalar field in $\AdS_{d+1}$ to the correlation function of four scalar
operators of arbitrary scale dimension in the boundary CFT. In the most
general case the amplitude is far more complicated than the calculation of gauge
boson exchange in [11], but there are simplifications if certain conditions
involving the dimensions $\Delta_i,\ \Delta$, and $d$ are satisfied.. 
We note here the role of the triangle inequality $\Delta <\Delta_1+\Delta_3$.
If satisfied, the final amplitude is as simple as for gauge exchange, and
the derivation could also be simplified by substituting (3.22) in (3.10) and
performing the $\gamma$ integral immediately, roughly as in [11]. It is also
striking that the triangle inequality is satisfied, up to marginal cases,
for chiral primary operators in the Type IIB $\AdS_5 \times S_5$ theory, and we
have proposed a particular interaction which simplifies the marginal cases in
the Appendix. When $\Delta >\Delta_1+\Delta_3$, the amplitude is
more complicated, and we have been able to show that its short-distance limit is
also different, contains $(\ln s)^2$ terms, and cannot obey the double OPE. We
observed that 3-point functions [20] also diverge in this case.

The major question still left open by this work is the question of logarithmic
singularities which have now appeared in all recent studies [9,8,11,14] of
4-point correlators. It is still the case that no complete correlator of
the $\AdS_5 \times S_5$ theory has been obtained, and we hope that a nearly
complete calculation of graviton exchange will achieve that goal [27]. 
If logarithms survive then the proposed interpretation [17] in terms of anomalous
dimensions of double-trace operators can be studied. The results for these
dimensions could then be new non-perturbative information from the AdS/CFT
correspondence.

\bigskip
\bigbreak

\centerline{{\bf A. APPENDIX}}

\bigskip

In this Appendix we discuss the marginal case $\Delta=\Delta_1+\Delta_3$ of
the triangle inequality $\Delta<\Delta_1+\Delta_3$. In particular we will
discuss the case $\Delta_1=\Delta_3=k$ and $\Delta=2k$ for $\AdS$ scalars
corresponding to chiral primary operators $\tr X^k$ of the $\AdS_5 \times S_5$
theory. Here, there is the somewhat paradoxical situation in that the
super Yang-Mills 3-point function $\langle \tr X^k (x_1) \tr X^{2k} (y)
\tr X^k(x_3)\rangle $ has a non-vanishing free-field amplitude shown to be equal
to the corresponding supergravity correlator [4] despite the fact that the
integrals for 3-point correlators computed in [20] diverge in this case. The
finite result in [4] comes from a zero in the cubic Lagrangian coupling which
multiplies the infinite value of the integral. Further, the simplifying
truncation for 4-point functions in the present paper fails for this case, and
the 4-point correlator appears to have $(\ln \ s)^2$ terms rather than the powers
expected in the double OPE.

\medskip

These paradoxes can be explained by the simple hypothesis that the actual
cubic couplings of the bulk fields $\phi_k$ and $\phi_{2k}$ have the 
particular form
$$
{\cal L}_k = \half c \big [
  D^mu\phi_{2k} \partial _mu (\phi_k^2) + m_{2k}^2 \phi_{2k} \phi_k^2 \big ],
\eqno  (A.1)
$$
where   $m_{2k}^2 = 2k(2k-d)$ is the mass squared of the heavier field,  
rather than the non-derivative coupling used in [4]. It is exactly for
this interaction that two divergent bulk integrals for the 3-point
function formally cancel by partial integration, leaving a boundary term, i.e.
$$
\eqalign{
V(x_1,x_3,y) & \equiv 
  c \int {d^{d+1} z \over z_0 ^{d+1}}
  \biggl \{ z_0 ^2  
  \partial _\mu \left ( {z_0 ^2 \over (z-x_1)^2 (z-x_3)^2} \right )^k
  \partial _\mu \left ( {z_0  \over (z-y)^2}               \right )^{2k}
 \cr
 & \qquad + {2k (2k-d) z_0 ^{4k} \over (z-x_1)^{2k} (z-x_3)^{2k} (z-y)^{4k}}
\biggr \}
\cr
       &  =\lim _{z_0 \to 0} c \int d^dz 
   {z_0^{2k-d+1} \over (z-x_1)^{2k} (z-x_3)^{2k}}
         \partial _0
\left ( { z_0 \over (z-y)^2} \right) ^{2k}. \cr}                
\eqno (A.2)
$$
Proceeding heuristically we note (3.20) that
$$ 
\lim _{z_0 \to 0}  \partial _0 \left ( {z_0\over (z-y)^2} \right )^{2k} \sim
{2k\over C_{2k}} z_0^{d-2k-1} \delta^{(d)} (z-y)
\eqno (A.3)
$$
where $C_{2k}$ is the normalization constant of (2.6). Thus the heuristic limit
of (A.2) is
$$
V(x_1,x_3,y) = {c\over C_{2k}}{ 1 \over (y-x_1)^{2k}}{1\over
(y-x_3)^{2k}}       
\eqno (A.4)
$$
which has the correct coordinate dependence. We cannot be certain that the
coefficient is correct because the situation is similar to that of 2-point
functions in [20] where it was shown that the heuristic limit gave an
incorrect coefficient and that a more careful calculation using Ward identities
or momentum space was required. This point needs further study.

\medskip

Next we discuss a 4-point function in which the interaction $L_k$ of (A.1)
appears at the $z$-vertex of Fig.~1 with the field of dimension $2k$
propagating to the $w$-vertex where there is a non-derivative interaction with
fields of dimension $\Delta_2$ and $\Delta_4$. We note that the partial
integration similar to that in (A.2) can be made with no residual boundary
term because $w_0 \not=0$. Instead we find $\big (\square -2k(2k-d)\big )
G_{2k}(u)= \delta^{d+1}(z-w)$ from the bulk-to-bulk propagator. Thus the
$z$-integral in the 4-point function can be done immediately leaving the
amplitude
$$
B(x_i)= \int {d^{d+1} w \over w_0^{d+1}}
 {w_0 ^{2k +\Delta _2 + \Delta _4} \over (w-x_1)^{2k} (w-x_3)^{2k}}
{1 \over (w-x_2)^{2\Delta _2} (w-x_4)^{2\Delta _4}}
\eqno (A.5)
$$
So the 4-point function is described by an effective contact interaction
for which a parametric integral representation [6,9] can be obtained.
We note that this amplitude has a finite limit as $x_3 \to x_1$ which has
the exact form of a 3-point function of operators of dimension $2k,
\Delta _2,\Delta_4$, and thus the coordinate dependence
$$
\lim _{x_3 \to x_1} B(x_i) = {c' \over (x_1-x_2)^{2k+\Delta2-\Delta_4} 
(x_1-x_4)^{2k+\Delta_4-\Delta_2} (x_2-x_4)^{\Delta_2+\Delta_4-2k}}       
\eqno (A.6)
$$
In the further limit $x_2 \to x_4$, we find exactly the expected contribution
to the double OPE of a primary operator of dimension $\Delta=2k$.

\medskip

Thus the hypothesis of an interaction of the form (A.1) appears to
resolve the paradoxes discussed above. It gives well defined 3- and 
4-point correlation functions whose coordinate dependence agrees with
expectations. It should be noted that for the case $\Delta>\Delta_1
+\Delta_3$, one can also define a combination of derivative and
non-derivative cubic interactions for which the divergent bulk integral
for 3-point functions can be formally replaced by a surface term.
In this case, however, the surface term is divergent, so it does not
seem possible to achieve well-defined 3-point correlators.

\medskip

There is still another issue to which the interaction (A.1) seems
relevant, namely the question of decoupling of fields in the graviton
multiplet from higher Kaluza-Klein modes in the $\AdS_5 \times S_5$ theory.
Since the correlator $\langle \tr X^2 \tr X^4 \tr X^2 \rangle$ is non-vanishing,
the theory certainly has a cubic coupling $\sim \phi_4 \phi_2 \phi _2$ when
expressed in terms of the fields that directly correspond to chiral primary
operators. However, one may try to redefine fields to eliminate the
coupling. For the Lagrangian
$$
{\cal L}= \half \big [D^\mu\phi_4 \partial _\mu \phi_4 +D^\mu \phi_2 \partial _mu
\phi_2 +  c''\phi_4 \phi_2 \phi _2 \big ]                 
\eqno (A.7)
$$
decoupling can be achieved by redefining $\phi_4=\Phi_4 +
1/2 c''\square ^{-1}\phi_2^2$ and decoupling by a non-local transformation
is generic for a combination of derivative and non-derivative
interactions. However, precisely for the interaction $L_k$ of (A.1)
with $k=2$, the decoupling transformation is local, viz. $\phi_4=
\Phi_4 - 1/2 c'\phi_2^2$. One should note that other interaction terms
such as $\phi_4^4$ or $\phi_4^2\phi_2^2$ can spoil local decoupling
unless they also appear in special combinations. So further study of
this and of the interactions of descendent fields is required before
the issue of local decoupling can be settled. The decoupling issue is
related to recent work [28] exploring     
classical solutions of the $\AdS_5 \times S_5$ theory which interpolate
between critical points. Non-local decoupling is apparently required
to justify the restriction to solutions involving only the metric
and scalars of the graviton multiplet and no Kaluza-Klein
excitations.\footnote{*}{We thank O. Aharony for this observation, and thank him
and also S. Gubser, K. Skenderis, P. van Nieuwenhuizen, and A. Zaffaroni 
for useful discussions of the decoupling question.}

\bigbreak

\centerline{{\bf ACKNOWLEDGMENTS}} 

\bigskip

It is a pleasure to acknowledge helpful conversations with Samir Mathur, Alec 
Matusis and Leonardo Rastelli. They are our collaborators on related projects
and have been generous with their advice for this one as well.  E. D. also
wishes to thank the members of the Institute for Theoretical Physics, Santa
Barbara for their hospitality while part of this work was completed. 

\bigskip
\bigbreak

\centerline{{\bf REFERENCES}}

\bigskip

\item{[1]} J.M.  Maldacena, ``The Large N Limit of Superconformal Field Theories
      and Supergravity, hep-th/9711200.

\item{[2]} S. Gubser, I.R. Klebanov and A.M. Polyakov, 
``Gauge Theory Correlators from Non-Critical String Theory",
Phys. Lett {\bf B428} (1998) 105; hep-th/9802109.

\item{[3]} E. Witten, ``Anti De Sitter Space And Holography'',
hep-th/9802150.

\item{[4]} S. Lee, S. Minwalla, M. Rangamani, and N. Seiberg,
  ``Three-Point Functions of Chiral Operators in D=4, $\N=4$ SYM at Large N'';
   hep-th/9806074.

\item{[5]} E. D'Hoker, D.Z. Freedman, and W. Skiba, ``Field Theory Tests for 
Correlators in the AdS/CFT Correspondence''; hep-th/9807098.

\item{[6]} W. Mueck and K.S. Viswanathan, ``Conformal Field Theory Correlators 
  from Classical Scalar Field Theory on $AdS_{d+1}$ ", 
  Phys. Rev. {\bf D58} (1998) 41901; hep-th/9804035.

\item{[7]} D.Z. Freedman, A. Matusis, S. D. Mathur, and L. Rastelli, as
discussed in Freedman's conference lecture at Strings '98, 
 http://www.itp.ucsb.edu/online/strings98/.

\item{[8]} H. Liu and A.A.  Tseytlin, ``On Four-point Functions in the
CFT/AdS Correspondence''; hep-th/9807097; G. Chalmers and K. Schalm, ``The
Large $N_c$ limit of Four-Point functions in N=4 super Yang Mills theory from
Anti-de Sitter Supergravity", hep-th/9810051.

\item{[9]} D.Z. Freedman, A. Matusis, S. D. Mathur, and L. Rastelli,
 ``Comments on 4-point functions in the CFT/AdS correspondence'',
  hep-th/9808006.

\item{[10]} S. Ferrara, conference lecture Strings '98, 
    {http://www.itp.ucsb.edu/online/strings98/}; S. Ferrara, R. Gatto, 
 A.F. Grillo, and G. Parisi, Nucl. Phys. {\bf B49} (1972) 77; Nuovo Cimento
 {\bf 26} (1975) 226; J.H. Brodie and M. Gutperle, ``String Corrections to four 
point functions in the AdS/CFT correspondence", hep-th/9808067.

\item{[11]} E. D'Hoker and D.Z. Freedman, ``Gauge Boson Exchange in
AdS$_{d+1}$", hep-th/9809179.

\item{[12]} G. Chalmers and K. Schalm, ``The large $N_c$ limit of Four point
functions in $N=4$ super Yang-Mills theory from Anti-de Sitter Supergravity",
hep-th/9810051.

\item{[13]} K. Intriligator, ``Bonus Symmetries of $\N=4$ super Yang-Mills
Correlation Functions via AdS duality", hep-th/9811047. 

\item{[14]} H. Liu, ``Scattering in Anti-de Sitter Space and Operator Product
Expansion", hep-th/9811152.

\item{[15]} F. Gonsalez-Rey, I. Park and K. Schalm, ``A Note on Four Point
Functions of Conformal Operators in $\N=4$ super Yang-Mills", hep-th/9811155.

\item{[16]} B. Eden, P.S. Howe, C. Schubert, E. Sokatchev and P.C. West,
``Four Point Functions in $\N=4$ super Yang-Mills theory at two loops",
hep-th/9811172.

\item{[17]} E. Witten, private communication.

\item{[18]} H.J. Kim, L.J. Romans and P. van Nieuwenhuizen, ``Mass spectrum of 
the chiral ten-dimensional $N=2$ supergravity on $S_5$", Phys. Rev. {\bf D32} 
(1985) 389.

\item{[19]} C. Fronsdal, Phys. Rev. {\bf D10} (1974) 589;
C.P. Burgess  and C. A. Lutken, Propagators and Effective Potentials in
Anti-de Sitter Space'', Nucl. Phys. {\bf B272} (1986) 661;
T.  Inami and H. Ooguri, ``One Loop Effective Potential in Anti-de Sitter 
Space'', Prog. Theo. Phys. {\bf 73} (1985) 1051;
C.J. Burges, D.Z. Freedman, S. Davis, and G.W. Gibbons, ``Supersymmetry
in Anti-de Sitter Space'', Ann. Phys. {\bf 167} (1986) 285. 

\item{[20]} D.Z. Freedman, A. Matusis, S. D. Mathur, and L. Rastelli,
``Correlation functions in the CFT(d)/AdS(d+1) correspondence'';
  hep-th/9804058.

\item{[21]} A. Erdelyi, {\it Bateman Manuscript Project, Higher Transcendental
Functions}, Krieger Publ. Comp. (1981) Vol I, page 111, (4).

\item{[22]} reference [21], page  64, (22).

\item{[23]} reference [21], page 110, (14).

\item{[24]} reference [21], page 110, (12).

\item{[25]} E. D'Hoker and D.H. Phong, ``Momentum Analyticity and Finiteness of
the 1-loop superstring Amplitude", Phys. Rev. Lett. {\bf 70} (1993) 3692;
``The Box Graph in Superstring Theory", Nucl. Phys. {\bf B440} (1995) 24.

\item{[26]} reference [21], page 108, (1).

\item{[27]} E. D'Hoker, D.Z. Freedman, S. Mathur, A. Matusis, L. Rastelli, to
appear.

\item{[28]}
 L.~Girardello, M.~Petrini, M.~Porrati, and A.~Zaffaroni, 
``Novel Local CFT and Exact Results on Perturbations of N=4
Super Yang Mills from AdS Dynamics'',  hep-th/9810126; 
 J.~Distler and F.~Zamora, ``Non-Supersymmetric Conformal Field Theories from
Stable Anti-de Sitter Spaces'', hep-th/9810206.


\bye